\documentclass[11pt,twoside]{article}
\usepackage{asp2010}
\usepackage{amssymb}
\usepackage{graphicx,mathptm} 
\newcommand{\be}{\begin{equation}}
\newcommand{\ee}{\end{equation}}
\newcommand{\bea}{\begin{eqnarray}}
\newcommand{\eea}{\end{eqnarray}}

\resetcounters

\begin{document}

\title{Cosmic Ray transport in MHD turbulence: large and small scale interactions}
\author{Huirong Yan$^1$, and A. Lazarian$^2$,
\affil{$^1$KIAA, Peking University, Beijing 100871, China}
\affil{$^2$Astronomy Department, University of Wisconsin, Madison, WI 53706, US}}

\begin{abstract}
Cosmic ray (CR) transport and acceleration is essential for many astrophysical problems, e.g., CMB foreground, ionization of molecular clouds and all high energy phenomena. Recent advances in MHD turbulence call for revisions in the paradigm of cosmic ray transport. We use the models of magnetohydrodynamic turbulence that were tested in numerical simulation, in which turbulence is injected at large scale and cascades to to small scales. We shall address the issue of the transport of CRs, both parallel and perpendicular to the magnetic field and show that the issue of cosmic ray subdiffusion is only important for restricted cases when the ambient turbulence is far from that suggested by numerical simulations. Moreover, on scales less than injection scale of turbulence, CRs's transport becomes super-diffusive. We also shall discuss the nonlinear growth of kinetic gyroresonance instability of cosmic rays induced by large scale compressible turbulence. This gyroresonance of cosmic rays on turbulence is demonstrated an important scattering mechanism in addition to direct interaction with the compressible turbulence. The feedback of the instability on large scale turbulence cannot be neglected, and should be included in future simulations.\end{abstract}

\section{Introduction}
The propagation and acceleration
of cosmic rays (CRs) is governed by their interactions
with magnetic fields. Astrophysical magnetic fields are turbulent and, 
therefore, the resonant and non-resonant (e.g. transient time damping, or TTD)
interaction of cosmic rays with MHD turbulence is the accepted
 principal mechanism to scatter and isotropize
cosmic rays \citep[see][]{Schlickeiser02}. In addition, efficient scattering is essential for the acceleration of cosmic rays. 
For instance, scattering of cosmic rays back into the shock is a
vital component of the first order Fermi acceleration \citep[see][]{Longairbook}. At the same time, stochastic acceleration by turbulence is 
entirely based on scattering. The dynamics of cosmic rays in MHD turbulence holds the key to all high energy astrophysics and related problems. 

We live in an exiting era when we are starting to test fundamental processes taking place at the beginning of the Universe, at the event horizon of black holes, when the nature of dark matter and dark energy is being probed etc. In the mean time, with the launching of the new facilities like Fermi, we have much more observational data available than ever before.  Using computers many researchers make sophisticated complex models to confront the observations in unprecedented details. This makes it urgent that we understand the key physical processes underlying astrophysical phenomena, can parameterize them and, if necessary, use as a subgrid input in our computer models.


At present, the propagation of the CRs is an advanced theory, which makes
use both of analytical studies and numerical simulations. However,
these advances have been done within the turbulence paradigm which
is being changed by the current research in the field.
Instead of the empirical 2D+slab model of turbulence, numerical
simulations suggest anisotropic Alfv\'enic modes (an analog of 2D, but not an
exact one, as the anisotropy changes with the scale involved) + fast modes
or/and slab modes \citep{CL02_PRL}.

At the same time, one should not disregard the possibilities of generation of additional perturbations on small scales by CR themselves. For instance,  the slab Alfv\'enic perturbation can be created, e.g., via streaming instability \citep[see][]{Wentzel74, Cesarsky80}. These perturbations are present for a range of CRs energies (e.g., $\lesssim 100$GeV in interstellar medium for the streaming instability) owing to non-linear damping 
arising from ambient turbulence (\citealp[][henceforth YL02, YL04]{YL02, YL04}, \citealp{FG04}).
Instabilities induced by anisotropic distribution of CRs were also suggested as a possibility to scatter CRs \citep[]{Lerche, Melrose74}. Gyroresonance instability induced by large scale turbulence compressions, can be an important feedback processes that creates slab modes to efficiently scatter CRs \citep{LB06, YL11}. 

Propagation of CRs perpendicular to mean magnetic field
is another important problem for which one needs to take into account both large and small scale interactions in tested models of turbulence. Indeed, if one takes only the diffusion along the magnetic field line and field line random walk \citep[FLRW][]{Jokipii1966, Jokipii_Parker1969, Forman1974}, compound (or subdiffusion) would arise. Whether the subdiffusion is realistic in fact depends on the models of turbulence chosen \citep{YL08}. In this paper we again seek the answer to  this questions within domain of numerically tested models of MHD turbulence.

In what follows, we discuss the cosmic ray transport in large scale turbulence in \S2. In \S3, we shall study the perpendicular transport of cosmic rays on both large and small scales. We shall also discuss the issue of super-diffusion and the applicability of sub-diffusion. In \S4, we address the issue of gyroresonance instability of CRs and its feedback on large scale compressible turbulence. Summary is provided in \S5.

\section{Cosmic ray scattering by compressible MHD turbulence}

As we mentioned earlier, numerical simulations of MHD turbulence supported the GS95 model of turbulence,
which does not have the "slab" Alfvenic modes that produced most of the scattering in the earlier models
of CR propagation. Can the turbulence that does not appeal to CRs back-reaction (see \S 4) produce 
efficient scattering? 

\begin{figure*} [h!t] 
{\includegraphics[width=0.32\textwidth,height=0.18\textheight]{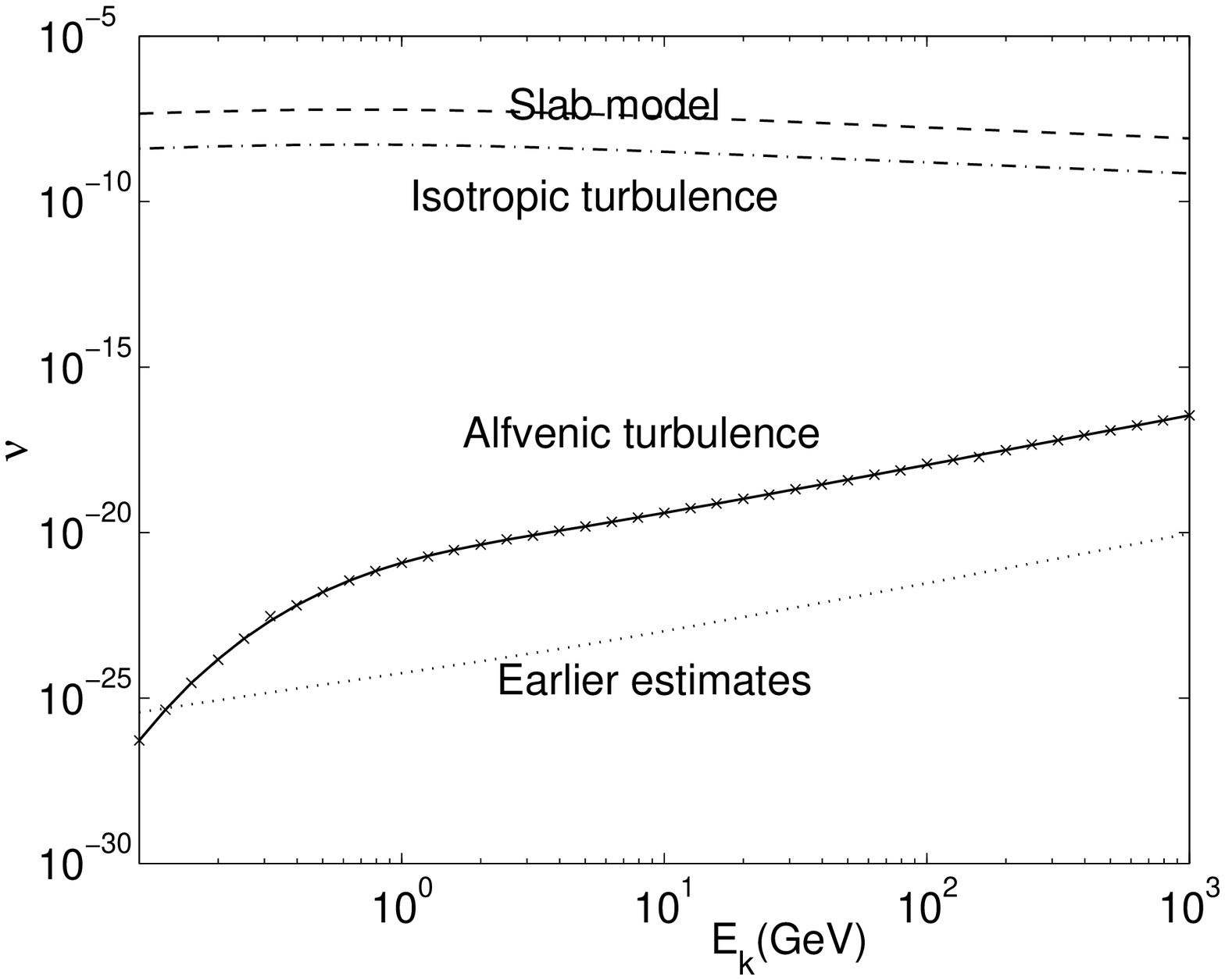} 
\includegraphics[width=0.32\textwidth,height=0.18\textheight]{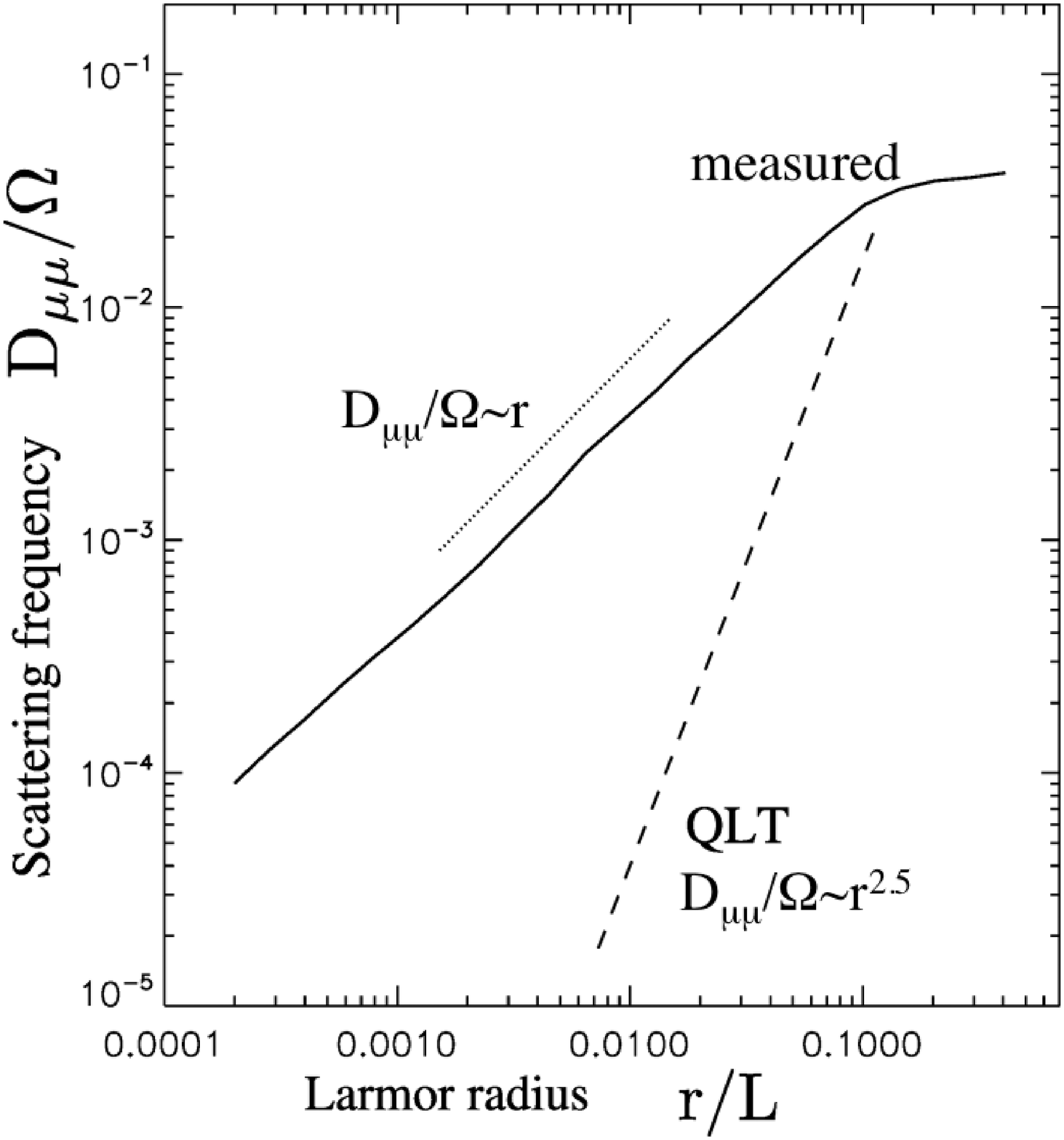}
\includegraphics[width=0.32\textwidth,height=0.18\textheight]{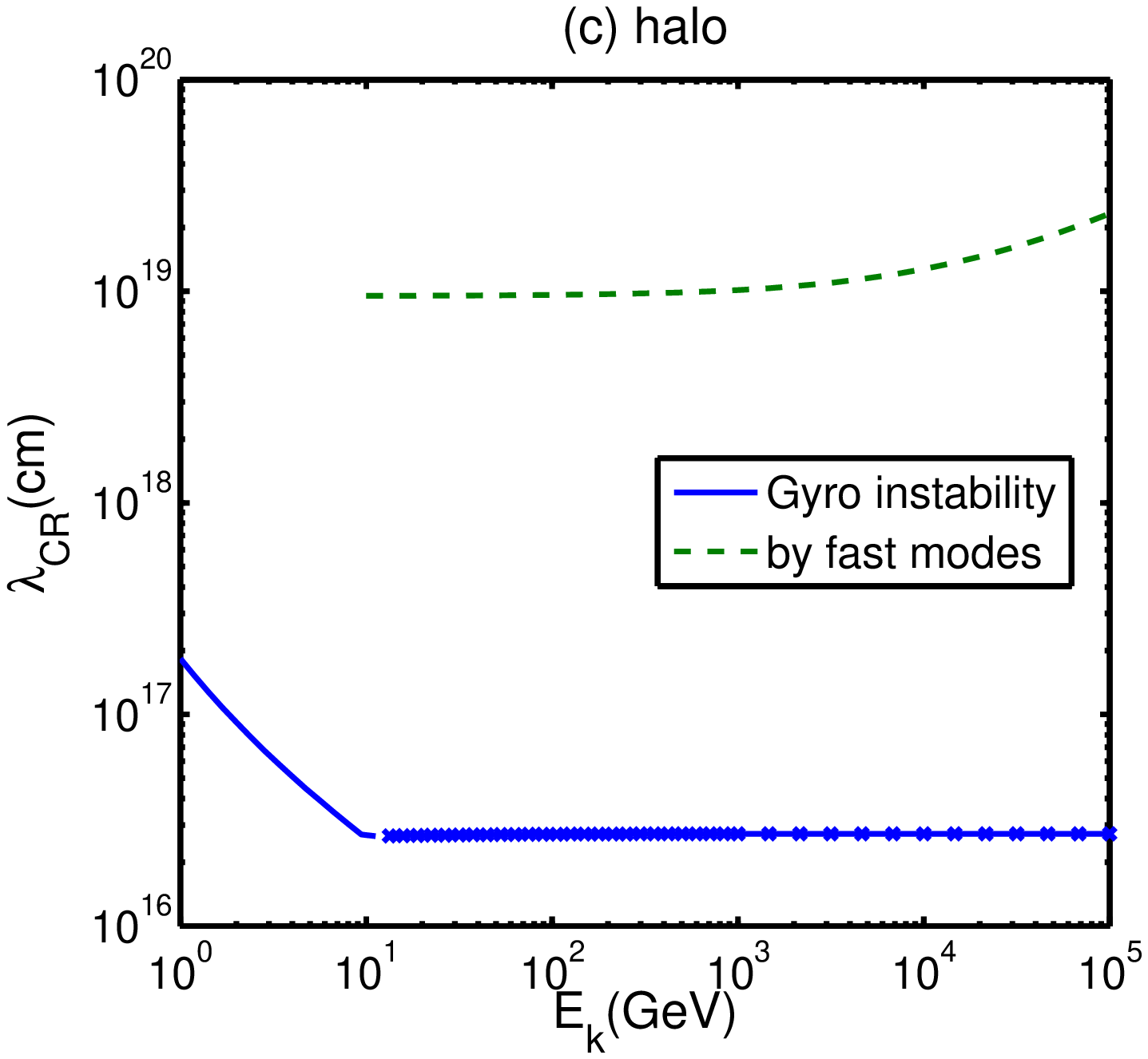}
} 
\caption{\small {\em Left:} rate of CR scattering by
Alfven waves versus CR energy.  The lines at the top of the figure are
the accepted estimates obtained for Kolmogorov turbulence. The dotted
curve is from \cite{Chandran00}. The analytical calculations are given
by the solid line with our numerical calculations given by
crosses. {\em Middle:} dimensionless CR scattering coefficient $D_{\mu
    \mu}/\Omega$ for the case of $\mu=0.71$ vs Larmor radius
  $r_L$ expressed in cube size units (solid line). We suppose
  that it is dominated by magnetic bottles formed by slow mode,
  this is why $D_{\mu \mu}$ (dimensional scattering frequency) is
  almost constant, i.e. independent on particle's energy. For
  comparison, we plot various theoretical predictions: QLT
  prediction for Alfven and slow mode (dashed) \citep*[from][]{BYL2011};  {\em Right:} mean free path of CRs vs. their energy. The scattering is dominated by direct interaction by fast modes for CRs with energies $\gtrsim$ a few tens GeV and through gyroresonance instability due to turbulence compression for lower energy CRs.}
\label{impl}
\end{figure*}

In the models of ISM turbulence \citep[]{Armstrong95, Mckee_Ostriker2007}, where the injection happens at large scale, 
fast modes were identified as a scattering agent for cosmic rays in interstellar medium \cite[]{YL02,YL04}.
These works made use of the quantitative description of turbulence
obtained in \cite{CL02_PRL}  to calculate
the scattering rate of cosmic rays. For instance, the scattering rate of relativistic protons by
Alfvenic turbulence was shown to be nearly $10^{10}$ times lower than
the generally accepted estimates obtained assuming the Kolmogorov
scaling of turbulence (see Fig.~\ref{impl}).  Although this estimate is $10^{4}$ times
larger than that obtained by \cite{Chandran00}, who employed
GS95 ideas of anisotropy, but lacked the quantitative
description of the eddies, it is clear that for most interstellar
circumstances the Alfvenic scattering is suppressed.  The low efficiency of scattering by
Alfvenic modes arise from high anisotropy of the modes at the scales of cosmic ray gyroradius.

Fast modes, however, are isotropic \citep{CL02_PRL}. Indeed they are subject to both
collisional and collisionless damping. The studies in \cite{YL02, YL04} demonstrated, nevertheless, that the scattering by fast modes dominates in most cases in spite of the damping\footnote{On the basis of weak turbulence theory, \cite{Chandran2005} has argued that high-frequency 
fast waves, which move mostly parallel to magnetic field, generate Alfven waves also moving mostly parallel to magnetic field. We expect
that the scattering by thus generated Alfven modes to be similar to the scattering by the fast modes created by them. Therefore
we expect that the simplified approach adopted in \cite{YL04} and the papers that followed to hold.}.
More recent studies of cosmic ray propagation and acceleration that explicitly appeal to the effect of
the fast modes include \citet{Cassano_Brunetti, Brunetti_Laz, YL08, YLP08}.
Incidentally, fast modes have been also identified as primary agents for the acceleration of charged dust particles \cite{YL03,YLD04}.

\section{Perpendicular transport}

In this section we deal with the diffusion perpendicular
to {\it mean} magnetic field. Both observation of Galactic CRs and solar wind indicate that the diffusion of CRs perpendicular to magnetic field is comparable to parallel diffusion \citep[]{Giacalone_Jok1999, Maclennan2001}. Why is that?

The perpendicular transport is slow if particles are restricted to the magnetic field 
lines and the transport is solely due to the random walk of field 
line wandering \citep[see][]{Kota_Jok2000}. 
In the three-dimensional turbulence, field lines are diverging away due to shearing by Alfv\'en modes \citep[see][]{LV99,Narayan_Medv,Lazarian06}.
 Since the Larmor radii of CRs are much larger than the minimum scale of eddies $l_{\bot, min}$, field lines within the CR Larmor orbit are effectively diverging away owing to shear by Alfv\'enic turbulence.
The cross-field transport  thus results from 
the deviations of field lines at small scales,
as well as field line random walk at large scale ($>{\rm min}[L/M^3_A,L]$), where $L, M_A$ are the injection scale and Alfv\'enic Mach number of the turbulence, respectively.

\subsection{Perpendicular diffusion on large scale}

{\it High $M_A$ turbulence}: High $M_A$ turbulence corresponds to the field that is easily bended by
hydrodynamic motions at the injection scale as the hydro energy at the
injection scale is much larger than the magnetic energy, i.e.
$\rho V_L^2\gg B^2$. In this case
magnetic field becomes dynamically important on a much smaller scale, i.e. the 
scale $l_A=L/M_A^3$ \citet[see][]{Lazarian06}. If the parallel mean free path of CRs $\lambda_\|\ll l_A$, the stiffness of B field is negligible so that the perpendicular diffusion coefficient is the same as the parallel one, i.e., $D_\bot=D_\|$. If $\lambda_\|\gg l_A$, the
 diffusion is controlled by the straightness of the field lines, and $
D_\bot=D_{\|}\approx 1/3l_Av,~~~M_A>1,~~~\lambda_{\|}>l_A.
\label{dbb}
$ The diffusion is isotropic if scales larger than $l_A$ are
concerned. In the opposite limit $\lambda_{\|}<l_A$, naturally, a result for
isotropic turbulence, namely, $D_{\bot}= D_{\|}\sim 1/3 \lambda_{\|} v$ holds. 

{\it Low $M_A$ turbulence}: For strong magnetic field, i.e. the field that cannot be easily bended at
the turbulence injection scale, individual magnetic field lines are aligned
with the mean magnetic field. The diffusion in this case is anisotropic.
If turbulence is injected at scale $L$ it stays 
weak for the scales larger than $LM_A^2$ and it is 
strong at smaller scales. Consider first the case of $\lambda_\|>L$.
The time of the individual step is $L/v_\|$, then $D_\perp\approx 1/3Lv M_A^4, ~~~M_A<1,~~~ \lambda_\|>L.$
This is similar to the case discussed in the FLRW model (Jokipii 1966). However, we obtain the dependence of $M_A^4$ instead of their $M_A^2$ scaling. In the opposite case of $\lambda_\|<L$, the perpendicular diffusion coefficient is $
D_{\bot}\approx D_{\|}M_A^4,
\label{diffx}$ which coincides with the result
obtained for the diffusion of thermal electrons in magnetized plasma \citep{Lazarian06}.

\subsection{Superdiffusion on small scales}

The diffusion of CR on the scales $\ll L$ is different and it is determined by how fast field lines are diverging away from each other. The mean deviation of a field in a distance $\delta x$ is proportional to $[\delta z]^{3/2}$  \citep{LV99, Lazarian06}, same as Richardson diffusion in the case of hydrodynamic turbulence \cite[see][]{Eyink2011}. Following the argument, we showed in \cite{YL08} that the cosmic ray perpendicular transport is superdiffusive. The reason is that there is no random walk on small scales up to the injection scale of strong MHD turbulence ($LM_A^2$ for $M_A < 1$
 and $l_A$ for $M_A>1$). 

\subsection{Is there subdiffusion?}
The diffusion coefficient $D_{\|}M_A^4$ we obtained in the case of $M_A<1$, means that the transport
perpendicular to the dynamically strong magnetic field is a diffusion, rather
than subdiffusion, as it was stated in a number of recent papers. This is also supported by test particle simulations \citep{BYL2011}. Let us
clarify this point by obtaining the necessary conditions for the subdiffusion
to take place.

The major implicit assumption in subdiffusion (or compound diffusion) is that the particles trace back 
their 
trajectories in x direction on the scale $\delta z$. When is it possible to talk about tracing particle trajectories back? In the case of random motions at a single scale {\it only}, the distance over 
which the particle
trajectories get uncorrelated is given by the \cite{RR1978}
model. Assuming that the damping scale of the turbulence  is larger
that the CR Larmor radius, this model, when generalized to anisotropic turbulence provides \citep{Narayan_Medv, Lazarian06} $L_{RR}=l_{\|, min}\ln(l_{\bot, min}/r_{Lar})$
where $l_{\|, min}$ is the parallel scale of the cut-off of turbulent motions, 
$l_{\bot, min}$ is the corresponding perpendicular scale, $r_{Lar}$ is the
CR Larmor radius. The assumption of $r_{Lar}<l_{\bot, min}$ can be
 valid, for instance, for Alfv\'enic motions in partially ionized gas.
However, it is easy to see that, even in this case,
 the corresponding scale is rather
small and therefore subdiffusion is not applicable for the transport
of particles in Alfv\'enic turbulence over scales $\gg l_{\|,min}$.

If  $r_{Lar}>l_{\bot, min}$, as it is a usual case for Alfv\'en motions in the
phase of ISM with the ionization larger than $\approx 93\%$, where the
Alfv\'enic motions go to the thermal particle gyroradius 
\citep[see estimates in][]{LG01, LVC04}, 
the subdiffusion of CR is not an applicable concept for Alfv\'enic turbulence.  
This does
not preclude subdiffusion from taking place
in particular models of magnetic perturbations,
e.g. in the slab model considered in \cite{Kota_Jok2000}, but we believe in the omnipresence of Alfv\'enic turbulence in interstellar gas \citep[see][]{Armstrong95}.

\section{Gyroresonance Instability of CRs in Compressible Turbulence}

Until recently, test particle approximation was assumed in most of earlier studies and no feedback of CRs is included apart from the streaming instability. Turbulence cascade is established from large scales and no feedback of CRs is included. This may not reflect the reality as we know the energy of CRs is comparable to that in turbulence and magnetic field \citep[see][]{Kulsrudbook}. It was suggested by \cite{LB06} that the gyroresonance instability of CRs can drain energy from the large scale turbulence and cause instability on small scales by the turbulence compression induced anisotropy on CRs. And the wave generated on the scales, in turn, provides additional scattering to CRs. In \cite{YL11}, we provided quantitative studies based on the nonlinear theory of the growth of the instability to take into account more accurately the feedback of the growing waves on the distributions of CRs.

In the presence of background compressible turbulence, the CR distribution is bound to be anisotropic because of the conservation of the first adiabatic invariant $\mu\equiv v_\bot^2/B$. Such anisotropic distribution is subjected to various instabilities. Waves are generated through the instabilities, enhancing the scattering rates of the particles, their distribution will be relaxed to the state of marginal state of instability even in the collisionless environment.  While the hydrodynamic instability requires certain threshold, the kinetic instability can grow very fast with small deviations from isotropy. Here, we focus on the gyroresonance instability. Both the qualitative and quantitative studies in \cite{YL11} show that the isotropization rate is roughly $
\tau^{-1}_{scatt} \sim \frac{\Gamma_{gr}\epsilon_N}{\beta_{CR} A}\label{nu_est}$, where $\Gamma_{cr}, \epsilon_N$ are the instability growth rate and the wave energy normalized by magnetic energy, respectively. $\beta_{CR}$ is the ratio of CR pressure to magnetic pressure, $A$ is the degree of anisotropy of the CR momentum distribution.

By balancing the rate of decrease in anisotropy due to scattering and the growth due to compression, one can get
\be
\epsilon_N\sim \frac{ \beta_{CR}\omega\delta v}{\Gamma_{gr} v_A },~~~\lambda_{CR}=r_p/\epsilon_N.
\label{epsilon_est},
\ee
where $v_A$ is the Alfv\'en speed, $\omega, \delta v$ are the wave frequency and amplitude at the scale that effectively compresses the magnetic field and create anisotropy in CRs' distribution \citep{YL11}.
  
\subsection{bottle-neck for the growth of the instability and feedback on turbulence}
\label{feedback}
The creation of the slab waves through the CR resonant instability is another channel to drain the energy of large scale turbulence. This process, on one hand, can damp the turbulence. On the other hand, it means that the growth rate is limited by the turbulence cascade. The energy growth rate  cannot be larger than the turbulence energy cascading rate, which is $1/2 \rho V_L^4/v_A/L$ for fast modes in low $\beta$ medium and $\rho v_A^3/l_A$ for slow modes in high $\beta$ medium. This places a constraint on the growth, thus the upper limit of wave energy is given by
\bea
\epsilon^u_N=\cases{ M_A^2 L_i/(L A)\gamma^{\alpha-1},& $\beta<1$ \cr
  L_i/(l_A A)\gamma^{\alpha-1}, & $\beta>1$, \cr}
\label{energy}
\eea
where $\gamma$ is the Lorentz factor and $L_i\simeq 6.4\times 10^{-7}(B/5{\rm \mu G})(10^-10{\rm cm}^3/n_{cr})$pc. The growth is induced by the compression at scales $\lesssim \lambda_{CR}$. Therefore, in the case that $\Gamma_{gr} \epsilon$ reaches the energy cascading rate, fast modes are damped at the corresponding maximum turbulence pumping scale $\lambda_{fb}=r_p/\epsilon_N$.  If $\lambda_{fb}$ is larger than the original damping scale $l_c$, then there is a feedback on the large scale compressible turbulence. This shows that test particle approach is not adequate and feedback should be included in future simulations.



\section{Summary}
\begin{itemize}
\item Compressible fast modes are most important for CR scattering. CR transport therefore varies from place to place.
\item Large scale mirror is essential for pitch angle scattering ( including 90 degree).
\item Subdiffusion does not happen in 3D turbulence .
\item Our results are tested using input from turbulence simulations.
\item Small scale slab waves are generated in compressible turbulence by gyroresonance instability, dominating the scattering of low energy CRs ($<100$GeV). Feedback of CRs on turbulence should be included in future simulations. 
\end{itemize}

\bibliography{yan}


\end{document}